\documentclass[letter]{aa} % for the letters 

\usepackage{graphicx}
\usepackage{txfonts}
\usepackage{natbib}
\bibpunct{(}{)}{;}{a}{}{,} % to follow the A&A style
\usepackage{booktabs}
\usepackage{multicol}
\usepackage{graphicx}
\usepackage{fancyhdr}
\usepackage{amsmath}	% Advanced maths commands
\usepackage{amssymb}	% Extra maths symbols
\usepackage{enumerate}
\usepackage{multirow}
\usepackage{mathtools}
\usepackage{longtable}
\usepackage{tabularx}
\usepackage{xcolor} % Colors used for comments commands
\usepackage{ulem} % Cancel text with horizontal bar
\usepackage{comment}
\usepackage{cuted}
\usepackage{soul}
\usepackage{hyperref}
\hypersetup{colorlinks=true,urlcolor=black, citecolor=blue, linkcolor=black}

\makeatletter
\renewcommand*\aa@pageof{, page \thepage{} of \pageref*{LastPage}}
\makeatother

\begin{document}

   \title{SDSS-V LVM: Detectability of Wolf-Rayet stars and their He\,II ionizing flux in low-metallicity environments}
    \titlerunning{Detectability of SMC WN3 stars and their He\,II ionizing flux}
   \subtitle{I. The weak-lined, early-type WN3 stars in the SMC}

   \author{
     G.\ Gonz\'alez-Tor\`a\inst{\ref{inst:ari}}
     \and
     A.\ A.\ C.\ Sander\inst{\ref{inst:ari},\ref{inst:iwr}}
     \and 
     E.\ Egorova\inst{\ref{inst:ari}}
     \and
     R.\ R.\ Lefever\inst{\ref{inst:ari}}
     \and
    V.\ Ramachandran\inst{\ref{inst:ari}}
    \and
    O.\ V.\ Egorov\inst{\ref{inst:ari}}
    \and
    J.\ Josiek\inst{\ref{inst:ari}}
    \and
    E.\ C.\ Sch{\"o}sser\inst{\ref{inst:ari}}
    \and
    M.\ Bernini-Peron\inst{\ref{inst:ari}}
    \and 
    K.\ Kreckel\inst{\ref{inst:ari}}
    \and
    A.\ Wofford\inst{\ref{inst:unamens}}
    \and
    O.\ G.\ Telford\inst{\ref{inst:utah},\ref{inst:prince},\ref{inst:car}}
    \and
    P.\ Senchyna\inst{\ref{inst:car}}
    \and
    C.\ Leitherer\inst{\ref{inst:stsci}}
    \and 
    F.-H.\ Liang\inst{\ref{inst:ari}}
    \and
    G.\ A.\ Blanc\inst{\ref{inst:car},\ref{inst:uchile}}
    \and
    N.\ Drory\inst{\ref{inst:mdout}}
    \and
    J.\ G.\ Fern\'andez-Trincado\inst{\ref{inst:ucn}}
    \and
    E.\ J. Johnston\inst{\ref{inst:udp}}
    \and
    A.\ J.\ Mejía-Narv{\'a}ez\inst{\ref{inst:uchile}}
    \and
    S.\ F.\ Sanchez\inst{\ref{inst:unamens},\ref{inst:iac},\ref{inst:ull}}
    }
    
    \institute{
   {Zentrum für Astronomie der Universität Heidelberg, Astronomisches Rechen-Institut, M{\"o}nchhofstr. 12-14, 69120 Heidelberg, Germany\label{inst:ari}},
   \email{gemma.gonzalez-tora@uni-heidelberg.de}
   \and
    {Universit\"at Heidelberg, Interdiszipli\"ares Zentrum f\"ur Wissenschaftliches Rechnen, 69120 Heidelberg, Germany\label{inst:iwr}}
   \and
   {Instituto de Astronom{\'i}a, Universidad Nacional Aut{\'o}noma de M{\'e}xico, A.P. 106, Ensenada 22800, BC, Mexico\label{inst:unamens}}
   \and
   {Department of Physics and Astronomy, University of Utah, 275 South University Street, Salt Lake City, UT 84112, USA\label{inst:utah}} 
   \and
   {Department of Astrophysical Sciences, Princeton University, 4 Ivy Lane, Princeton, NJ 08544, USA\label{inst:prince}}
   \and 
   {The Observatories of the Carnegie Institution for Science, 813 Santa Barbara Street, Pasadena, CA 91101, USA\label{inst:car}}
   \and
   {Space Telescope Science Institute, 3700 San Martin Dr, Baltimore, MD 21218, USA\label{inst:stsci}}
   \and
   {McDonald Observatory, The University of Texas at Austin, 1 University Station, Austin, TX 78712-0259, USA\label{inst:mdout}}
   \and
   {Universidad Cat\'olica del Norte, Instituto de Astronom\'ia, Av. Angamos 0610, Antofagasta, Chile\label{inst:ucn}}
   \and
   {Univ. Diego Portales, Inst. de Estudios Astrofísicos, Fac. de Ingeniería y Ciencias, Av. Ejército Libertador 441, Santiago, Chile\label{inst:udp}}
   \and
   {Departamento de Astronom\'ia, Universidad de Chile, Camino del Observatorio 1515, Las Condes, Santiago, Chile\label{inst:uchile}}
   \and  
   {Instituto de Astrof\'\i sica de Canarias, La Laguna, Tenerife, E-38200, Spain \label{inst:iac}}
   \and 
   {Departamento de Astrof\'\i sica, Universidad de La Laguna, Spain\label{inst:ull}}
    }

   \date{Received \today; accepted YYYY}

\abstract{

The Small Magellanic Cloud (SMC) is the nearest low-metallicity dwarf galaxy. Its proximity and low reddening has enabled us to detect its Wolf-Rayet (WR) star population with 12 known objects. Quantitative spectroscopy of the stars revealed half of these WR stars to be strong sources of \ion{He}{ii} ionizing flux, but the average metallicity of the SMC is below where WR bumps are usually detected in integrated galaxy spectra showing nebular \ion{He}{ii} emission. Utilizing the Local Volume Mapper (LVM) integral-field spectroscopic survey, we investigate regions around the six SMC WN3h stars, whose winds are optically thin at $\geq$$\,54\,$eV, allowing these energetic photons to escape. Focusing on \ion{He}{ii}$\,4686\,$\AA\,, we show that the broad stellar wind component, the strongest optical diagnostic of WN3h stars, is diluted within 24\,pc in the integrated spectra, making such WR stars hard to detect in unresolved low-metallicity regions.
In addition, we compare the \ion{He}{ii} ionizing flux from LVM with the values inferred from the stellar atmosphere code PoWR and find that in nearly all cases, the stars emit more than enough hard ionizing photons to explain the observed \ion{He}{ii} nebular emission. We conclude that early-type WN stars with comparably thin winds are viable sources to produce the observed \ion{He}{ii} ionizing flux in low-metallicity galaxies. The easy dilution of the stellar signatures can explain the rareness of WR bump detections at $12 + \log\,\text{O}/\text{H} < 8.0$, while at the same time providing major candidates for the observed excess of nebular \ion{He}{ii} emission. This is challenging for population synthesis models across all redshifts as the evolutionary path towards this observed WR population at low metallicity remains enigmatic.
} 
  

   \maketitle

\section{Introduction}

The presence of nebular \ion{He}{ii} recombination emission in both high-redshift galaxies and nearby, metal-poor dwarf galaxies implies the existence of sources of high energy ionizing photons ($\geq54$eV). Recently, metal-poor, star-forming galaxies were discovered by JWST \citep{Pontoppidan+2022} observations at high redshift \citep[$z\geq7$, e.g.,][]{Schaerer+2022,Arellano-Cordova+2022,Robertson+2023,Trussler+2023}, showing some puzzling results such as unusual chemical abundance patterns, in particular strong nitrogen enrichment \citep[e.g.,][]{Cameron+2023} and high ionization \citep[e.g.,][]{Calabro+2024}. Both aspects are apparently correlated, with \citet{Topping+2025} recently reporting about a galaxy at $z = 7.04$ with nebular N\,IV] and C\,IV detections having equivalent widths larger than $5$ and $10\,\AA$, respectively.

Massive stars ($M_{\mathrm{init}}>8M_{\odot}$) spend most of their life in a hot stage ($T_{\mathrm{eff}}>10$\,kK), with their maximum of spectral energy distribution in the ultraviolet (UV). These stars release photons with high enough energy to ionize the surrounding interstellar medium, in particular producing \ion{H}{i} ($\lambda<912\,\AA$), \ion{He}{i} ($\lambda<504\,\AA$) and \ion{He}{ii} ($\lambda<228\,\AA$). 
Ionizing fluxes for a certain ion, $Q_{\mathrm{edge}}$, are expressed as
    $Q_{\mathrm{edge}}:= 4 \pi R_\ast^2 \int_{\nu_{\mathrm{edge}}}^{\infty} \frac{F_{\nu}}{h\nu} \,\mathrm{d}\nu$,
with $\nu_{\mathrm{edge}}$ the integrated frequency beyond reaching the sufficient energy to ionize the element and $F_{\nu}$ the stellar flux. The \ion{He}{ii} ionizing flux, $Q_{\mathrm{HeII}}$, requires very hot temperatures and optically thin stellar winds at $\geq54$eV, being strongly dependent on metallicity \citep[e.g.,][]{Guseva2000,Schaerer2003}. Classical, i.e., He-burning, Wolf-Rayet (WR) stars have suitable intrinsic temperatures, but their strong, often optically thick winds prohibit the escape of \ion{He}{ii} ionizing photons \citep[e.g.,][]{Smith2002,Sander2022}. The winds of WR stars with otherwise similar parameters become more optically thin with lower metallicity, eventually making them transparent to hard ionizing flux \citep{Sander2023,Sander+2025} before losing the spectral appearance that defines them as WR stars. Early-type WR stars
in low-metallicity environments with thinner winds ($\dot{M}_\text{t} \lesssim 10^{-4.5} M_\odot\,\mathrm{yr}^{-1}$) are therefore significant contributors of $Q_{\mathrm{HeII}}$ \citep[e.g.,][]{CrowtherHadfield2006,Hainich2015,Sander2023,Sander+2025}. However, many star-forming galaxies with strong \ion{He}{ii} nebular emission do not show WR features \citep[e.g,][]{Shirazi2012,Senchyna2017}.

The Small Magellanic Cloud (SMC) is the only very nearby \citep[$d=62.44 \pm 0.47$ kpc,][]{Graczyk2020} galaxy with a significant population of resolvable hot, massive stars and a significantly sub-solar metallicity \citep[$Z \approx 0.2\,Z_\odot$,][]{Bouret2003,Ramachandran2019}. 12 massive WR stars are known in the SMC \citep{Westerlund1964, Smith1968, Sanduleak1968, Sanduleak1969, Breysacher1978, Azzopardi1979, Morgan1991, Massey2001, Massey2003, Neugent2018}. Six of them are early-type WN (WNE) stars -- all WN3h -- with optically thin enough winds at $\geq54$eV to be strong $Q_{\mathrm{HeII}}$-contributors \citep{Hainich2015,Shenar2016}, including the binary system AB\,7. In addition, there is a WN3 star in the higher-order multiple system AB\,6, but it does not seem to produce \ion{He}{ii} ionizing flux  \citep{Shenar2018}. 

The detection of single WR stars without any stellar or compact companion at low $Z$ -- in particular in the SMC -- provides an ongoing challenge \citep[e.g.,][]{Shenar2020,Schootemeijer2024}. With decreasing wind mass loss at lower metallicity in the regime of hot stars, intrinsic self-stripping of the outer layers, also known as the ``Conti scenario'' \citep{Conti1975}, becomes less feasible with no current standard evolution model predicting the present population of weak-lined WN2 and WN3 stars in the Magellanic Clouds as discussed in more detail in \citet{Schootemeijer2024} and \citet{Sander+2025}.
Binary mergers \citep[e.g.,][]{Vanbeveren1997, Schuermann+2024} and chemically homogeneous evolution \citep[e.g.,][]{Martins2009,Szecsi2015,Boco+2025} offer potential formation alternatives, but so far fail to reproduce the characteristics of the observed single WRs with high $Q_{\mathrm{HeII}}$ and thinner winds. This in turn affects population synthesis models \citep[e.g., Starbust99, BPASS, C\&B;][]{Leitherer+2014,Hawcroft+2025,Plat+2019} which are missing a source of strong kinetic and ionizing feedback at low $Z$. As the broad, stellar emission lines of the SMC WN3 stars are rather weak, the non-detection of WR bumps in integrated light might not necessarily imply the absence of such WR stars. In this work, we use the new opportunity provided by the SDSS-V Local Volume Mapper \citep[LVM,][]{Drory2024, Kollmeier2025} to study and quantify the dilution of the strongest optical emission line of the SMC WN3 stars, \ion{He}{ii}\,4686\,\AA. The extended narrow, nebular \ion{He}{ii}\,4686\,\AA\  enables direct measurements of $Q_{\mathrm{HeII}}$. It is detected with LVM in numerous SMC regions, and associated with various types of sources (Egorova et al., in prep.) including all 6 WR stars selected for this study (see Table~\ref{tab:qheii}). As stellar sources of $Q_{\mathrm{HeII}}$ are rare, the WN3 stars are in most cases the dominating source \citep{Sander2022}, providing us with a rare opportunity to directly compare stellar and nebular determinations.

In Appendix\,\ref{sec:data}, we describe the observations and employed data. In Sect.\,\ref{sec:dilution} we analyze the dilution of the WR emission, before estimating and comparing \ion{He}{ii} ionizing fluxes in Sect.\,\ref{sec:fluxes}. We summarize our findings and conclusions in Sect.\,\ref{sec:summary}.

\section{WR line dilution and nebular emission} \label{sec:dilution}

\begin{figure}[t]
      \centering
      \includegraphics[width=.9\linewidth]{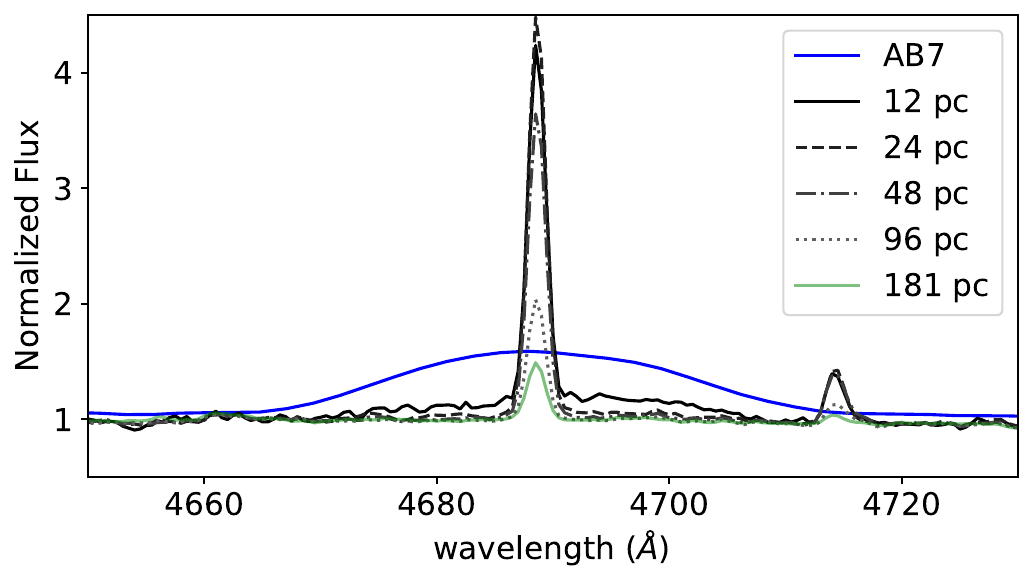}
      \caption{Spectra with normalized continuum around \ion{He}{ii} $4686\,$\AA\ for the WN3h star SMC AB\,7 (blue) and regions with different apertures from the LVM data, black for the smallest, lighter green for wider apertures.
      }
      \label{fig:ab7dilution}
\end{figure}

Utilizing the observations and stellar atmosphere models described in Appendix\,\ref{sec:data}, we can compare the individual spectra and ionizing fluxes with the results from integrated-light data.
In Figure\,\ref{fig:ab7dilution} we show the normalized stellar spectrum for AB\,7 around \ion{He}{ii} $\lambda\,4686\,$\AA\ and the integrated light from LVM. The broad, stellar emission line is always diluted in the LVM data, vanishing quickly with larger aperture. Intensity maps for all apertures are shown in Figures \,\ref{fig:ab179map} and \ref{fig:ab101112map}. Appendix\,\ref{ap:heii} gives the full set of spectra and  explanations on target differences.

Only four of our six targets (AB\,1, AB\,7, AB\,9, and AB\,12) yield a noticeable imprint of the broad stellar wind line for the 40$\arcsec$ aperture. Beside a dilution due to nebular continuum, these stars also differ in their luminosities \citep[$\log L/L_\odot \gtrsim 5.9$,][]{Hainich2015,Shenar2016} and their stellar neighborhood. In contrast, a target like AB\,10 ($\log L/L_\odot = 5.56$) has a more typical luminosity for the bulk of the known, resolved classical WR population and does not leave any stellar imprint in the 40$\arcsec$ aperture.
AB\,11 ($\log L/L_\odot = 5.9$) is marginally detected, but likely due to a lack of brighter nearby stars which would further dilute the WR emission lines. Within 40$\arcsec$, AB\,11 is clearly the star with the brightest B and V magnitudes. This is not the case for AB\,10, which has nearby sources more than an order of magnitude brighter in both bands, e.g., AzV\,7 (B0).

The blend of the broad \ion{He}{ii} $\lambda\,4686\,$\AA\ emission line in integrated galaxy spectra is called the ``blue WR bump'', mainly present in star-forming galaxies and first detected in the blue compact dwarf galaxy He 2-10 by \citet{Allen+1976}. 
The broad \ion{He}{ii} $\lambda\,4686\,$\AA\ can be generated by both WN and WC stars, but the latter typically also have strong \ion{C}{iv} $\lambda\,5808\,$\AA\ emission which give rise to a second bump denoted as ``yellow'' or ''red''. In the case of the SMC, no WC stars are known, but there is a WO4 star in the wing with significant \ion{C}{iv} emission \citep[e.g.,][]{Bartzakos+2001,Shenar2016} which would dominate any yellow bump if noticeable within an integrated SMC spectrum \citep{Crowther+2023}. In principle, \ion{C}{iv} $\lambda\,5808\,$\AA\ emission can also be seen in WN stars (though to a much smaller extend), but it is completely absent in the population of SMC WN3h stars due to the combination of a low carbon abundance and high temperature (and thus high excitation status) in the stellar atmosphere. As such, these stars can at best contribute to a blue bump. 

The strength of WR bumps is a popular method to estimate the number of WRs in an unresolved region or galaxy. The LVM data enables us to check popular estimates from \citet{Lopez-Sanchez2010} and \citet{Crowther+2023} for the blue bump. As we explain in Appendix\,\ref{ap:wrbump}, WR populations containing mainly weak-lined WNE stars are a challenge for the approach by \citet{Lopez-Sanchez2010}. The more specific relation from \citet{Crowther+2023} for SMC WNE stars gives a reasonable number on the order of unity if a WR bump is detected, but requires to make a prior informed decision. Appendix\,\ref{ap:ews} shows the equivalent width (EW) ratios for the extended LVM apertures with respect to the stellar EW, showing that the stellar component is diluted even at 12 pc. At $d>24$pc, the ratio is $< 0.025$ in all cases, making the WR undetectable. 

Overall, we can conclude that the broad emission of any SMC WN3h star is completely diluted (i.e., undetectable) if the flux is integrated over a region with a diameter $>$\,24\,pc. This value, which can become even lower in cases of lower WR luminosities or crowding with other bright stars, is considerably smaller than the typical resolution achievable with an IFU in many nearby galaxies \citep[$\sim$50\,pc, e.g., in PHANGS-MUSE or DGIS, see][]{Emsellem+2022,Li+2025}. Thus, weak-lined WNE stars are easy to hide in integrated light and have been missed in even less resolved galaxy surveys like CALIFA and MaNGA \citep{Miralles-Caballero+2016,Liang+2020}.

\section{He II ionizing fluxes} \label{sec:fluxes}

The quantitative spectroscopic analysis with PoWR obtained by \citet{Hainich2015} for SMC AB\,1, 9, 10, 11 and 12, and \citet{Shenar2016} for SMC AB\,7 provides us with a prediction for the intrinsic $\log\,Q_{\text{HeII}}$ of the SMC WN3 stars, which we use as a baseline to compare with our nebular line measurements from the LVM data. These models do not include a consistent solution of the wind hydrodynamics for the velocity field \citep{Sander2017,Sander2020a}, which is intrinsically complicated for the weak-lined WN3 stars due to radiatively-driven turbulence arising near the wind onset in this regime  \citep{Moens2022}. Based on preliminary test calculations, we do not expect drastically effects for the predicted $Q_{\text{HeII}}$. For WN2 stars, where the wind onset is not affected by the same issues, dynamically-consistent models can yield higher $Q_{\text{HeII}}$ of up to 0.8\,dex \citep{Sander+2025}, but mainly due to a temperature revision in the hydrodynamic analysis, which we do not expect for the WN3 stars.

The lower panels of the Figures in Appendix~\ref{ap:heii} show the LVM flux-calibrated spectra focusing on the \ion{He}{ii}$\,4686\,$\AA\, line for different apertures. 
To estimate $\log\,Q_{\text{HeII}}$, we use the formula from \citet{Kehrig2015}:
\begin{equation}
    \text{Q}(\text{HeII})=L_{\text{HeII}}/[j(\lambda 4686)/\alpha_{\text{B}}(\text{HeII})]\approx L_{\text{HeII}}/3.66\times10^{-13} ,
\end{equation}
where $j(\lambda 4686)$ is the photon energy at 4686$\,\AA\,$ and $\alpha_{\text{B}}(\text{HeII})$ the recombination rate coefficient assuming case B recombination and $T_\text{e}\sim2\times10^{4}$K \citep{Oserbrock2006}. We have calculated the $L_{\text{HeII}}$ with the dereddend LVM spectra, using the same extinction as \citet{Hainich2015,Shenar2016} and the same reddening law \citep[][plus a small MW foreground]{Gordon2003}. The resulting $\log\,Q_{\text{HeII}}$ are shown in Tab.~\ref{tab:qheii}. 
\begin{table*}
\caption{Fundamental parameters of the SMC WN3h stars and their \ion{He}{ii} ionizing fluxes}

\label{tab:qheii}
\small
\centering

\begin{tabular}{l c c c c c c c c c c c c}
      \hline \hline
       Target\rule[1.0em]{0em}{0em} 
             & \multicolumn{5}{c}{Stellar parameters\tablefootmark{(a)}} & & \multicolumn{6}{c}{measured $\log\,Q_{\mathrm{HeII}}$ (phot.\,s$^{-1}$) from LVM} \\
              
             & $T_\text{eff}(\tau = 2/3)$  & $\log L$ & $\log \dot{M}_\text{t}$\tablefootmark{(b)}  & $\log\,q_{\mathrm{HeII}}$\tablefootmark{(c)}  &   $\log\,Q_{\mathrm{HeII,star}}$ &  &  & $40\arcsec$ & $80\arcsec$ & $160\arcsec$ & $320\arcsec$ & $600\arcsec$  \\
             & (kK) & ($L_\odot$) & ($M_\odot\,\mathrm{yr}^{-1}$) & (phot.\,s$^{-1}$\,cm$^{-2}$)  & (phot.\,s$^{-1}$) &  &       &  12\,pc        &     24\,pc    &     48\,pc   &     96\,pc    &     181\,pc      \\\hline
         AB1\rule[1.1em]{0em}{0em} 
             &  79 & 6.07 & $-5.36$ & 23.53 & 47.68   &     &       &  46.445 & 46.540 & 46.913 & 47.055 & 47.073\\\smallskip
  &     &      &      &       & \multicolumn{2}{r}{$Q_{\mathrm{HeII,LVM}}$/$Q_{\mathrm{HeII,star}}$:}  &  & 0.0582 & 0.0725 & 0.1709 & 0.2374 & 0.2473\\

         AB7 &  98 & 6.10 & $-4.81$ & 24.67 & 48.50   &     &       &  47.933 & 48.352 & 48.546 & 48.583 & 48.695\\\smallskip
  &     &      &      &       & \multicolumn{2}{r}{$Q_{\mathrm{HeII,LVM}}$/$Q_{\mathrm{HeII,star}}$:}  &  & 0.2710 & 0.7116 & 1.1124 & 1.2118 & 1.5654\\

         AB9 &  99 & 6.05 & $-5.44$ & 24.61 & 48.49   &     &       &  46.253 & 46.728 & 47.185 & 47.808 & 48.102\\\smallskip
  &     &      &      &       & \multicolumn{2}{r}{$Q_{\mathrm{HeII,LVM}}$/$Q_{\mathrm{HeII,star}}$:}  &  & 0.0058 & 0.0173 & 0.0496 & 0.2080 & 0.4096\\

         AB10 & 98 & 5.65 & $-5.38$ & 24.62 & 48.10   &     &       &  47.448 & 47.385 & 47.345 & 47.374 & 47.549\\\smallskip
  &     &      &      &       & \multicolumn{2}{r}{$Q_{\mathrm{HeII,LVM}}$/$Q_{\mathrm{HeII,star}}$:}  &  & 0.2229 & 0.1927 & 0.1759 & 0.1880 & 0.2814\\
         AB11 &  88 & 5.85 & $-5.49$ & 23.90 & 47.88   &     &       &  45.794 & 46.029 & 46.502 & 47.068 & 47.427\\\smallskip
  &     &      &      &       & \multicolumn{2}{r}{$Q_{\mathrm{HeII,LVM}}$/$Q_{\mathrm{HeII,star}}$:}  &  & 0.0082 & 0.0141 & 0.0419 & 0.1540 & 0.3522\\

         AB12 & 112 & 5.90 & $-5.47$ & 25.04 & 48.57   &     &       &  46.447 & 46.786 & 47.236 & 47.605 & 48.456\\\smallskip
  &     &      &      &       & \multicolumn{2}{r}{$Q_{\mathrm{HeII,LVM}}$/$Q_{\mathrm{HeII,star}}$:}  &  & 0.0075 & 0.0165 & 0.0464 & 0.1085 & 0.7690\\
\hline
\end{tabular}
\tablefoot{
\tablefoottext{a}{Taken from \citet{Hainich2015} and \citet{Shenar2016}}
\tablefoottext{b}{Defined by \citet{Graefener2013}: $\dot{M}_\text{t} \propto \dot{M}\,\sqrt{D}\,\varv_\infty^{-1} L^{-3/4}$ with $D$ denoting the clumping factor and $\varv_\infty$ the terminal wind velocity}
\tablefoottext{c}{$q = \frac{Q}{4 \pi R_\ast^2}$, where $R_{\ast}$ is the extended stellar radius given by the PoWR models.}
}
\end{table*}

Using the ionizing flux ratios from Tab.~\ref{tab:qheii}, we compare the nebular diagnostics with the PoWR analysis results and obtain $Q_{\mathrm{HeII,LVM}}$/$Q_{\mathrm{HeII,star}}<1$ for all cases except the apertures of AB\,7 with $\mathrm{r}>80\,\arcsec$. This means that the star alone could produce more than enough photons to ionize the region comprised on the whole fiber apertures. For the AB\,7 apertures with $\mathrm{r}>80\,\arcsec$, the blending with other nearby sources from NGC\,395 or SNR\,E0102-72 in larger apertures could provide an extra source of ionization.

\section{Summary and Conclusions}\label{sec:summary}
 
The availability of LVM provides a unique opportunity to compare resolved stars with integrated light. Using the region around the \ion{He}{ii} $\lambda 4686\,\AA$ line, we study the imprint of the strongest optical emission line of SMC WN3 stars and of their \ion{He}{ii} ionizing fluxes in the LVM data, taking different apertures.
The broad stellar wind component is always diluted and becomes undetectable in the LVM data when integrating over more than 24\,pc in diameter.
For the case of AB\,1, 9, 10, 11, and 12, the \ion{He}{ii} ionizing fluxes inferred from detailed atmosphere modelling are at least $\sim$1.5 times larger than the values measured from the nebular emission. This means that a significant number of ionizing photons from the WN3 targets escape from the immediate environment. In contrast, the larger apertures for AB\,7 show an example where the WR alone can produce only up to $\sim$60\% of the the observed \ion{He}{ii} ionizing flux and other, less extreme nearby objects are also contributing to the observed nebular line emission.

While other possible sources of \ion{He}{ii} ionizing flux like massive stellar compact end products such as high-mass X-ray binaries \citep[HMXB, e.g.,][]{Plat+2019,Schaerer2019}, intermediate-mass stripped stars \citep{Goetberg+2019,Goetberg+2023}, and ultra-luminous X-ray \citep[ULX, e.g.,][]{Schaerer2019,Mayya2023} sources so far could not provide a promising explanation to the significant presence of nebular \ion{He}{ii} emission in high-$z$ galaxies and metal poor local galaxies \citep[e.g.,][Egorova et al., in prep.]{Senchyna2020,Saxena2020,Wofford+2021},
our results show that early-type WN stars with optically thin winds at $\geq54$eV are rather easy to ``hide'', even when they are significantly away from larger clusters, despite their strong contribution of \ion{He}{ii} ionizing photons. While likely not being the exclusive origin of nebular \ion{He}{ii}, so far current population synthesis models do not reach such an observed WR stage already at SMC metallicities, thereby lacking a potential major contributor. Updated treatments will have major impacts on stellar population predictions and cosmological simulations of galaxy evolution.

\begin{acknowledgements}
We thank and E.\ Tarantino for useful discussions. GGT is supported by the Federal Ministry for Economic Affairs and Climate Action (BMWK) via the Deutsches Zentrum f\"ur Luft- und Raumfahrt (DLR) grant 50 OR 2503 (PI Sander). AACS, RRL, and VR acknowledge support by the Deutsche Forschungsgemeinschaft (DFG) in the form of an Emmy Noether Research Group -- Project-ID 445674056 (SA4064/1-1, PI Sander). JJ is supported by the DFG under Project-ID 496854903 (SA4064/2-1, PI Sander). ECS is supported by the BMWK via the DLR grant 50 OR 2306 (PI Ramachandran/Sander).
This project was co-funded by the European Union (Project 101183150 - OCEANS). 
EE, KK, and FHL acknowledge funding from the European Research Council’s starting grant ERC StG-101077573 (`ISM-METALS', PI Kreckel). OE acknowledges funding
from the DFG under Project-ID 541068876 (PI Egorov). This work was performed in part at Aspen Center for Physics, which is supported by National Science Foundation grant PHY-2210452. SFS thanks the support by UNAM PASPA – DGAPA, the SECIHTI CBF-2025-I-236 project, and the Spanish Ministry of Science and Innovation
(MICINN), project PID2019-107408GB-C43 (ESTALLIDOS). GAB is supported by the ANID Basal project FB210003.

Funding for the Sloan Digital Sky Survey V has been provided by the Alfred P. Sloan Foundation, the Heising-Simons Foundation, the National Science Foundation, and the Participating Institutions. SDSS acknowledges support and resources from the Center for High-Performance Computing at the University of Utah. SDSS telescopes are located at Apache Point Observatory, funded by the Astrophysical Research Consortium and operated by New Mexico State University, and at Las Campanas Observatory, operated by the Carnegie Institution for Science. The SDSS web site is \url{www.sdss.org}.

SDSS is managed by the Astrophysical Research Consortium for the Participating Institutions of the SDSS Collaboration, including the Carnegie Institution for Science, Chilean National Time Allocation Committee (CNTAC) ratified researchers, Caltech, the Gotham Participation Group, Harvard University, Heidelberg University, The Flatiron Institute, The Johns Hopkins University, L'Ecole polytechnique f\'{e}d\'{e}rale de Lausanne (EPFL), Leibniz-Institut f\"{u}r Astrophysik Potsdam (AIP), Max-Planck-Institut f\"{u}r Astronomie (MPIA Heidelberg), Max-Planck-Institut f\"{u}r Extraterrestrische Physik (MPE), Nanjing University, National Astronomical Observatories of China (NAOC), New Mexico State University, The Ohio State University, Pennsylvania State University, Smithsonian Astrophysical Observatory, Space Telescope Science Institute (STScI), the Stellar Astrophysics Participation Group, Universidad Nacional Aut\'{o}noma de M\'{e}xico, University of Arizona, University of Colorado Boulder, University of Illinois at Urbana-Champaign, University of Toronto, University of Utah, University of Virginia, Yale University, and Yunnan University.  

\end{acknowledgements}

\bibliographystyle{aa}
\bibliography{references}

\newpage
\appendix
\section{Observations and atmosphere analyses} \label{sec:data}

The goal of the LVM survey is to map the Milky Way plane, Magellanic Clouds, and a sample of southern, nearby galaxies. The detailed description of the LVM telescope system, science motivation, and technical strategy are given in \cite{Herbst+2024,Drory2024,Sanchez+2025}. The LVM instrument is a wide field integral field unit (IFU) telescope built and operated at Las Campanas Observatory (LCO) by SDSS-V \citep{Kollmeier2025}.
The LVM IFU has a field of view of 0.5\,$\deg$\ with 1801 hexagonally packed fibers of 35.3$\arcsec$ apertures, with spectral coverage $3600-9800\,\AA$, and spectral resolution R$\sim4000$.
All observations of the Magellanic Clouds have a nominal total exposure time of 8100\,s, consisting of nine individual 900\,s-exposures, dithered following a nine-point grid pattern. 

\begin{figure*}[ht!]
      \centering
      \includegraphics[width=1.\linewidth]{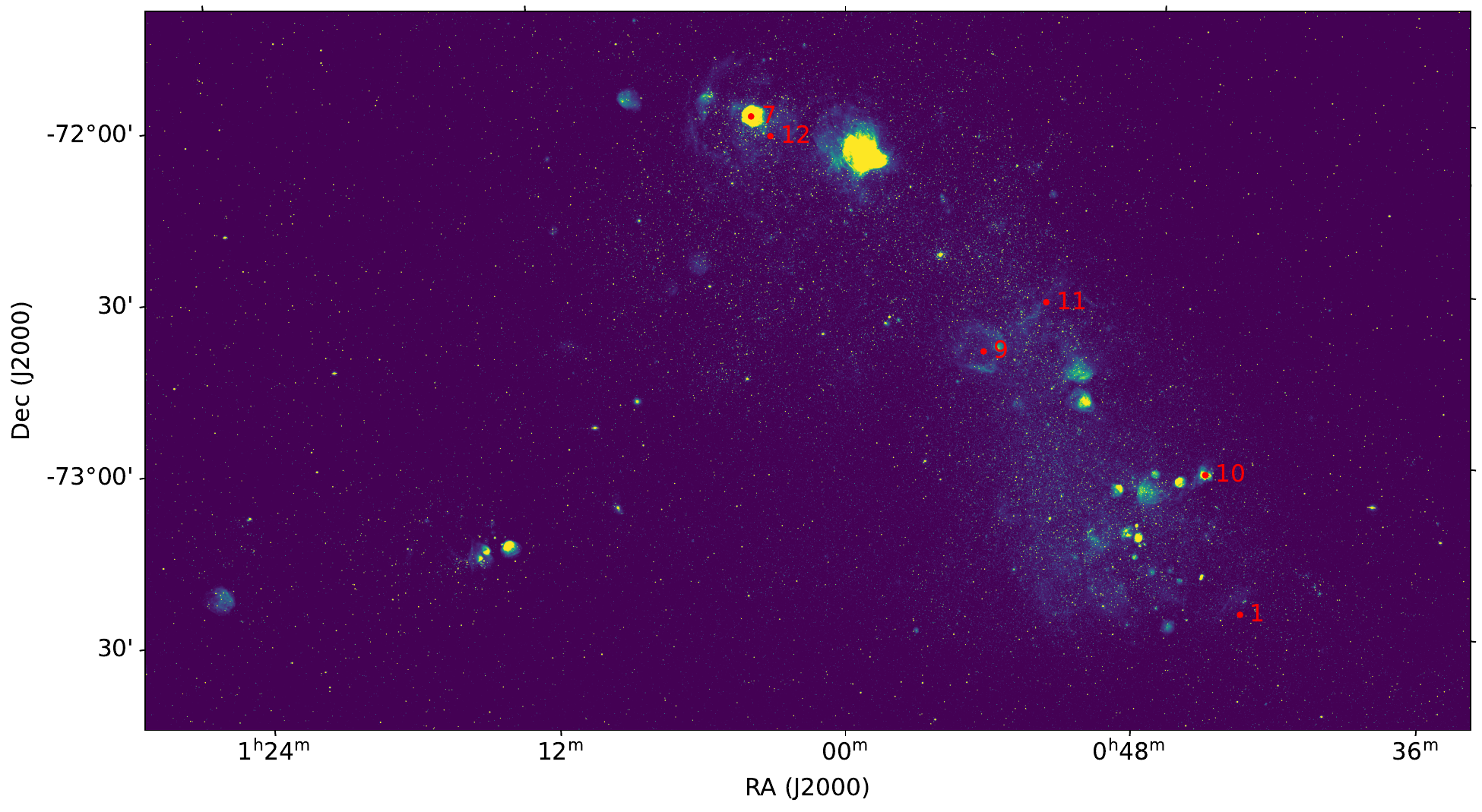}
      \caption{ Position of the WN stars studied in this work, using the identifiers by their number as SMC AB\#. The background image shows the \ion{O}{iii} nebular emission from MCELS \citep{Smith2005}.  }
      \label{fig:smc}
\end{figure*}

In this work, we focus on the six WN3h SMC stars, whose surrounding nebular emission is covered by the LVM fibers. Figure~\ref{fig:smc} shows the location of the targets in the SMC. One more WR system, the WO+O binary AB8 in the SMC Wing, has been identified as a considerable source of $Q_{\text{HeII}}$ from quantitative spectroscopy \citep[$\sim$$5 \cdot 10^{47}\,\mathrm{s}^{-1}$,][]{Shenar2016,Ramachandran2019}. However, the LVM has not covered the Wing region yet and we thus focus on the already fully covered sample of early-type WN stars in the SMC in this work.

\begin{figure*}[h!]
      \centering
      \includegraphics[width=.33\linewidth]{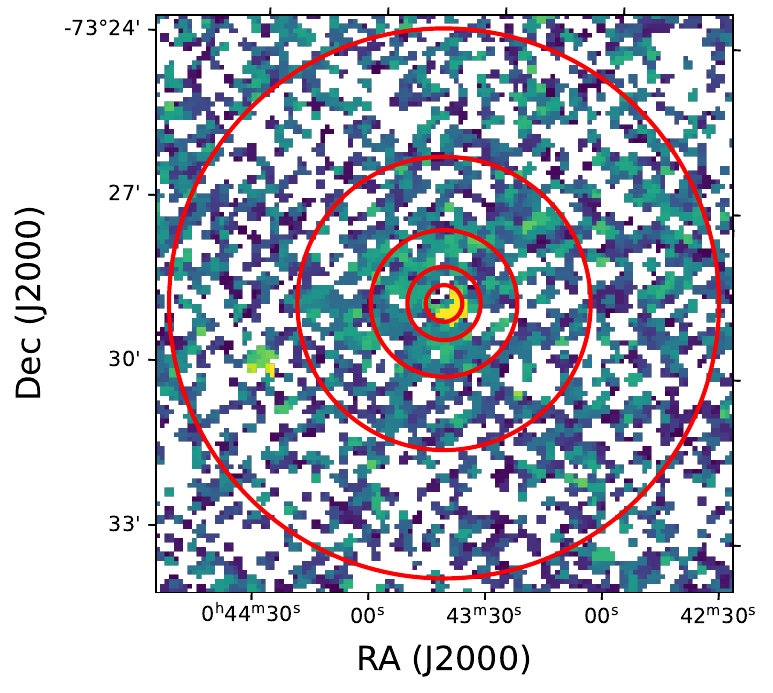}\hfill
      \includegraphics[width=.33\linewidth]{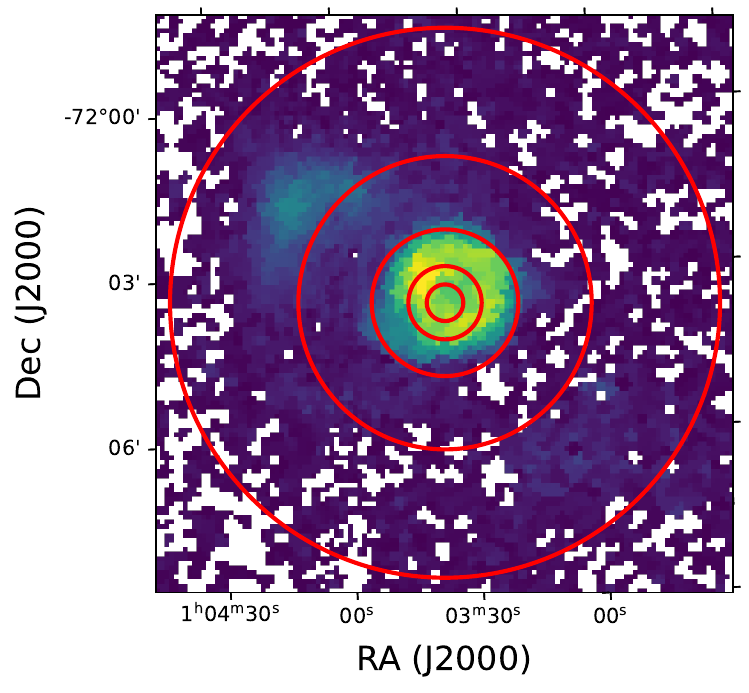}\hfill
      \includegraphics[width=.33\linewidth]{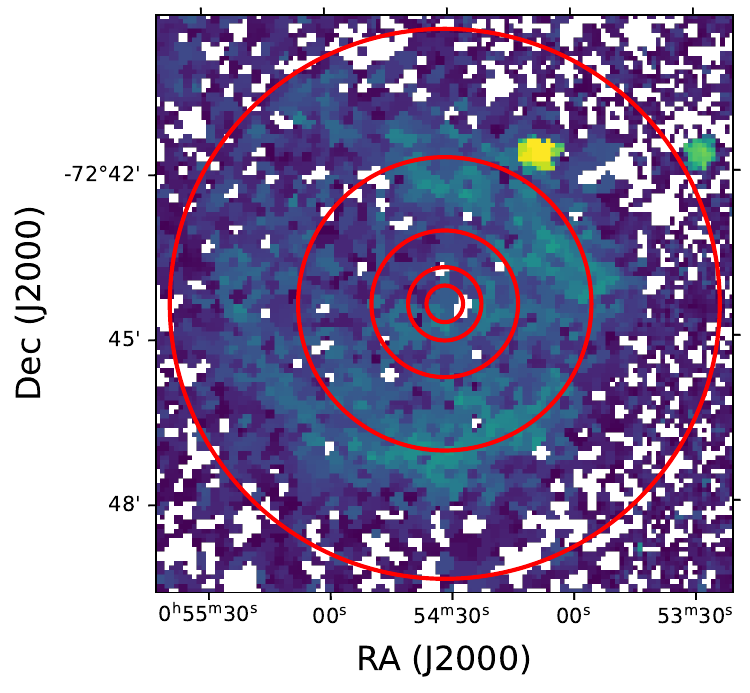}
      \caption{ The \ion{He}{ii} $ - 4686\,$\AA\ intensity map centered at SMC AB1 (left), SMC AB7 (center), and SMC AB9 (right) with the corresponding apertures in blue of 40$\arcsec$ (12\,pc), 80$\arcsec$ (24\,pc), 160$\arcsec$ (48\,pc), 320$\arcsec$ (96\,pc) and 600$\arcsec$ (181\,pc) of diameter.
      }
      \label{fig:ab179map}
\end{figure*}

\begin{figure*}[h!]
      \centering
      \includegraphics[width=.33\linewidth]{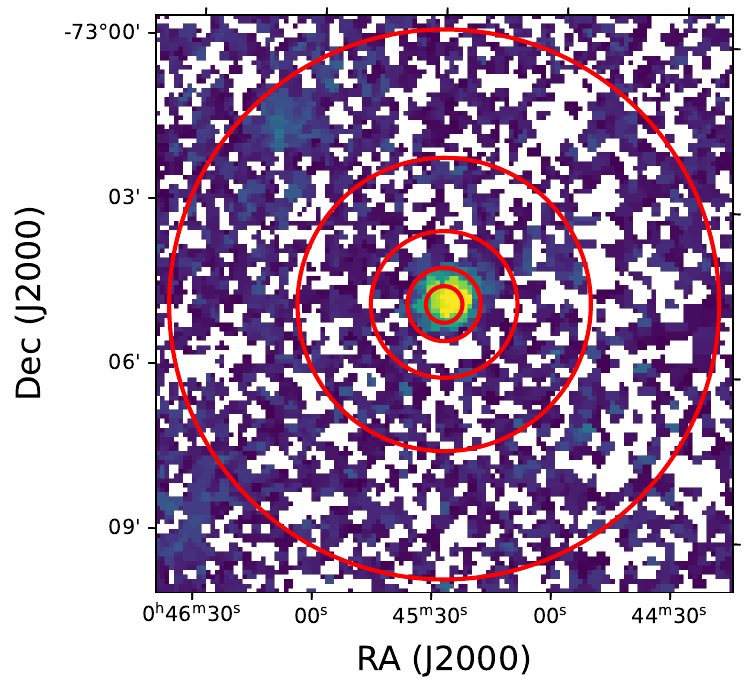}\hfill
      \includegraphics[width=.33\linewidth]{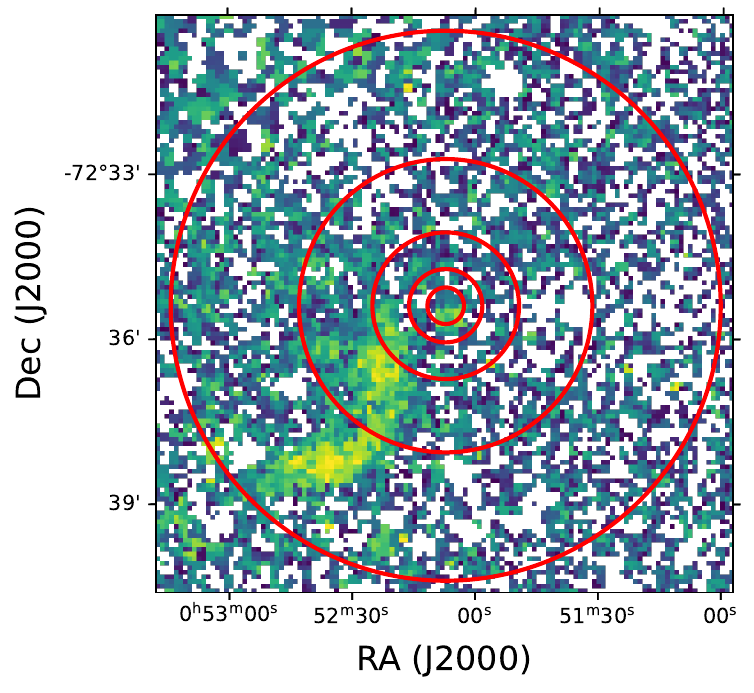}\hfill
      \includegraphics[width=.33\linewidth]{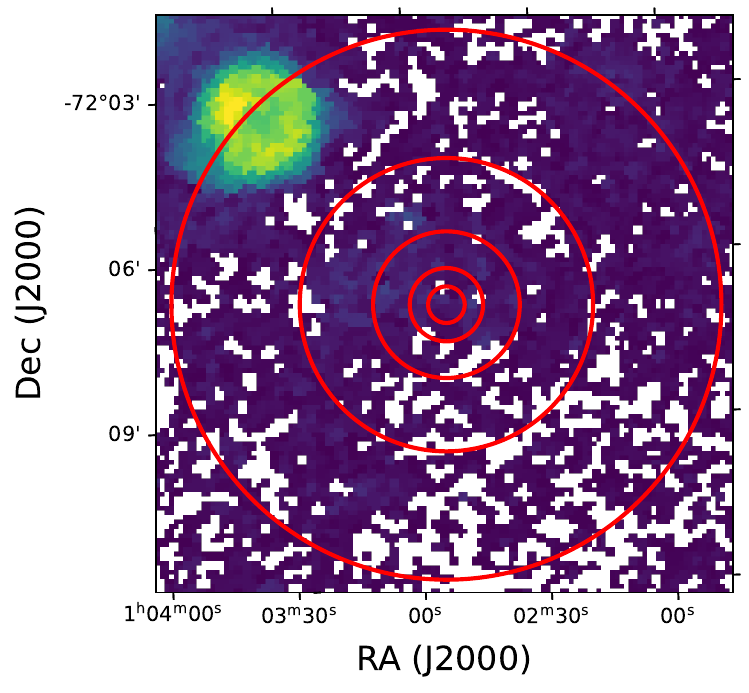}
      \caption{Same as Fig.~\ref{fig:ab179map} but for SMC AB10 (left), SMC AB11 (center), and SMC AB12 (right). 
      }
      \label{fig:ab101112map}
  \end{figure*}

For all the targets we have plotted the integrated spectra from apertures with 40$\arcsec$, 80$\arcsec$, 160$\arcsec$, 320$\arcsec$, and 600$\arcsec$ diameter. The final spectra were obtained by integrating the spectra from all the fibers that fall inside the given aperture weighted by their effective area (accounting for overlaps by assigning shared pixels fractionally) normalized by the total fiber area. 
In Figs.\,\ref{fig:ab179map} and \ref{fig:ab101112map}, we present intensity maps in the \ion{He}{ii} $ - 4686\,$\AA\ extracted from the LVM data for the different WN3 stars studied in this work. We further overplot the different apertures we apply for the spectral extraction and comparisons. For some targets, such as AB\,9, AB\,11, or AB\,12 (cf.\ \ion{He}{ii} maps in Figs.\,\ref{fig:ab179map} and \ref{fig:ab101112map}), the largest apertures contain nearby sources also contributing to \ion{He}{ii} emission, e.g., the sgB[e] LHA 115-S 18 for AB\,9 \citep{Zickgraf+1989} or AB\,7 falling in the largest aperture around AB\,12. These cases illustrate how source blending is an issue in the case of unresolved populations.

For the comparison with the stellar spectra of the WN3 stars, we use optical slit spectroscopy ($3700-6830\,\AA$) from \citet{Foellmi2003, Foellmi2004}.
The properties of the SMC WR stars were analyzed in \citet{Hainich2015} and \citet{Shenar2016,Shenar2018} using the stellar atmosphere code PoWR  \citep{Graefener+2002,HamannGraefener2003,Sander2015}.  

\section{The \ion{He}{ii} $ - 4686\,$\AA\ line }\label{ap:heii}
\begin{figure}[h!]
      \centering
      \includegraphics[width=1.\linewidth]{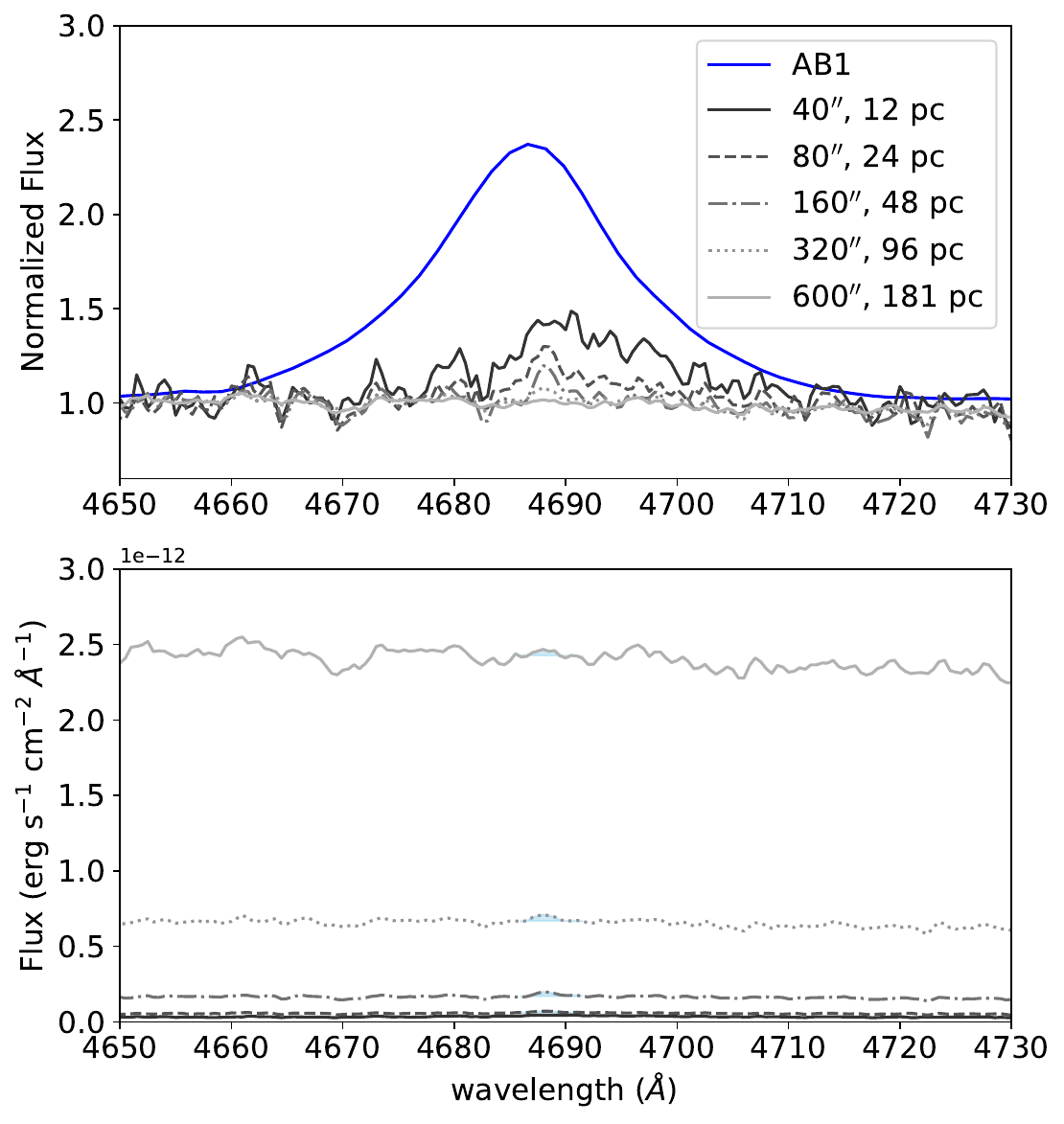}
      \caption{ The \ion{He}{ii} $ - 4686\,$\AA\ line profile for the SMC AB1 region. \textit{Upper panel: }The normalized flux for the \ion{He}{ii} $ - 4686\,$\AA\ line profile for the SMC AB1 target, in blue for the data of the star \citep[from][]{Foellmi2003}, black for the smallest fiber aperture of LVM (40$\arcsec$), lighter gray for wider apertures. \textit{Lower panel: } The calibrated flux for the \ion{He}{ii} $ - 4686\,$\AA\ line profile for the SMC AB1 nebular region, black for the smallest fiber aperture of LVM (40$\arcsec$), lighter gray for wider apertures. In light blue we show the regions selected of the \ion{He}{ii} that contribute to the nebular component. 
      }
      \label{fig:ab1heii}
  \end{figure}

\begin{figure}[h!]
      \centering
      \includegraphics[width=1.\linewidth]{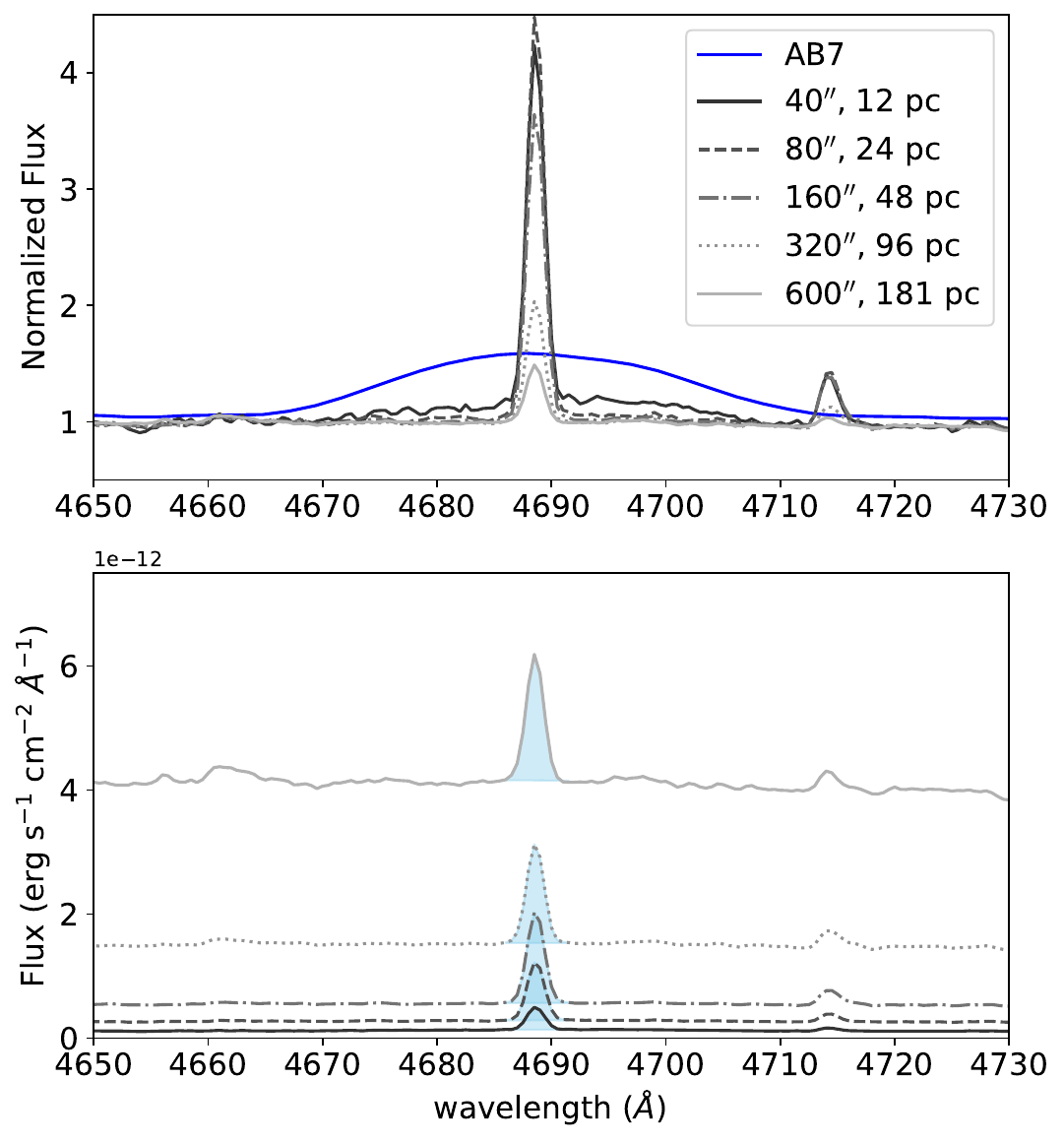}
      \caption{ Same as Fig.~\ref{fig:ab1heii} but for SMC AB7. 
      }
      \label{fig:ab7heii}
  \end{figure}

\begin{figure}[h!]
      \centering
      \includegraphics[width=1.\linewidth]{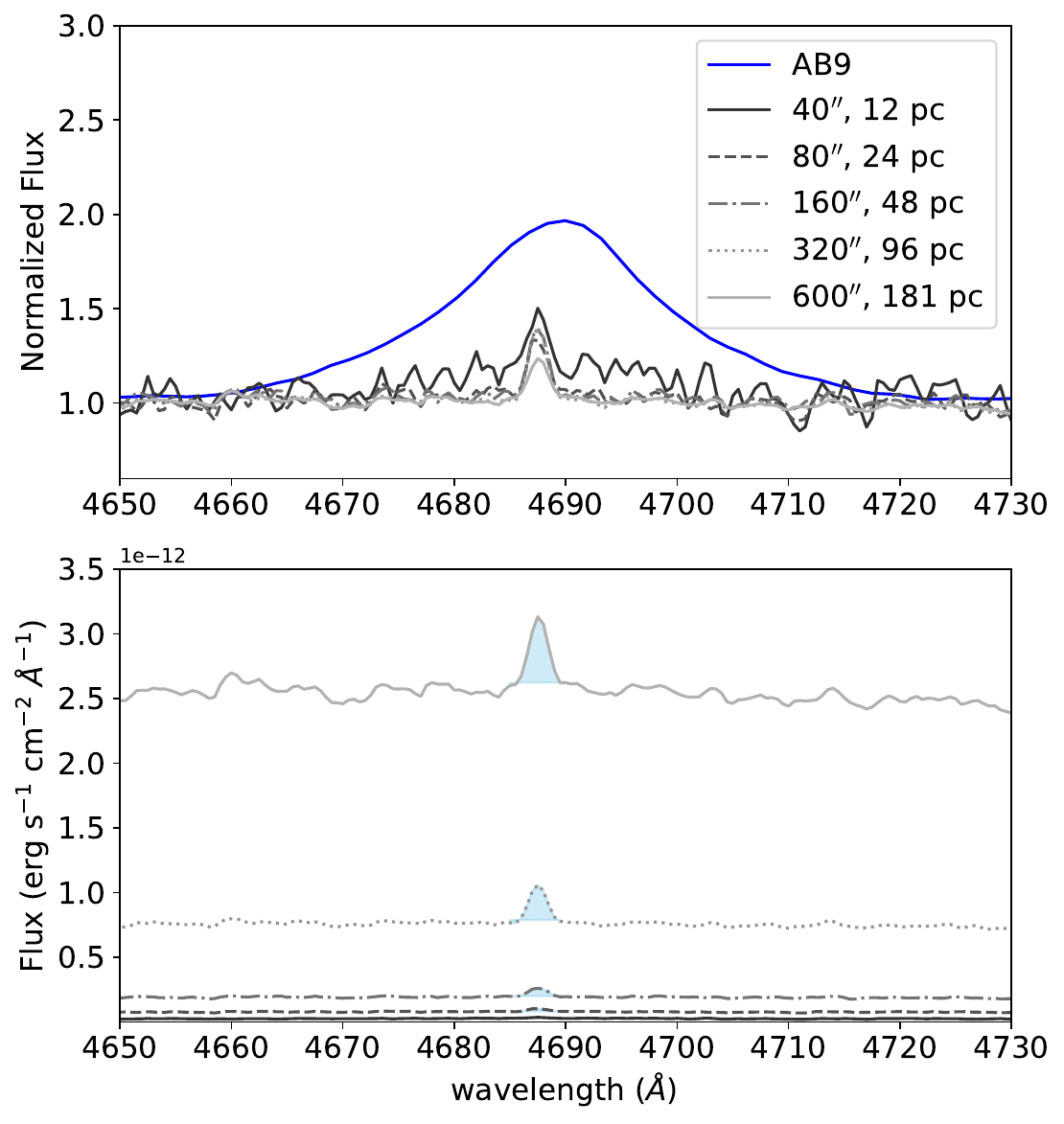}
      \caption{ Same as Fig.~\ref{fig:ab1heii} but for SMC AB9. 
      }
      \label{fig:ab9heii}
  \end{figure}

\begin{figure}[h!]
      \centering
      \includegraphics[width=1.\linewidth]{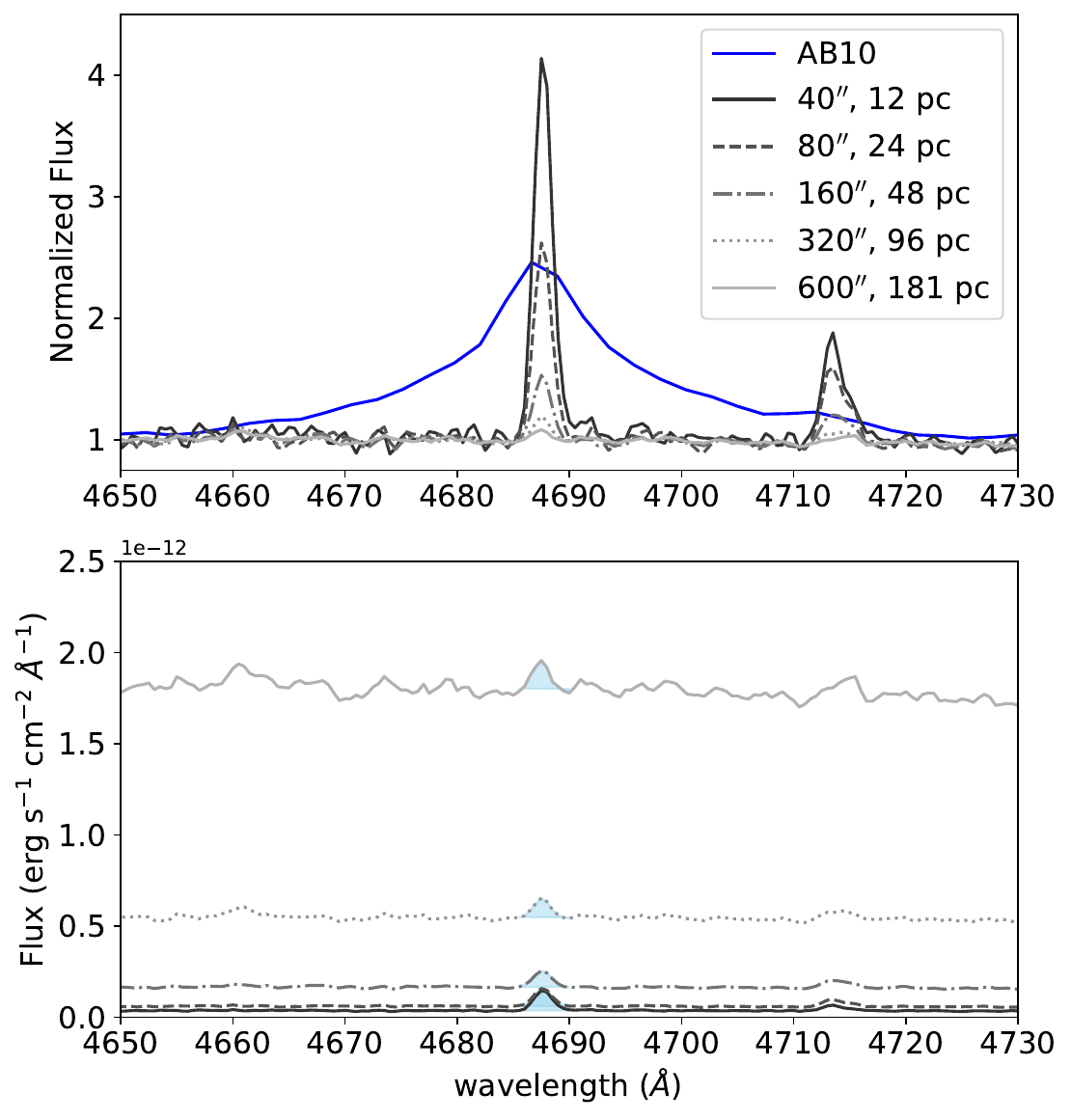}
      \caption{ Same as Fig.~\ref{fig:ab1heii} but for SMC AB10. 
      }
      \label{fig:ab10heii}
  \end{figure}

\begin{figure}[h!]
      \centering
      \includegraphics[width=1.\linewidth]{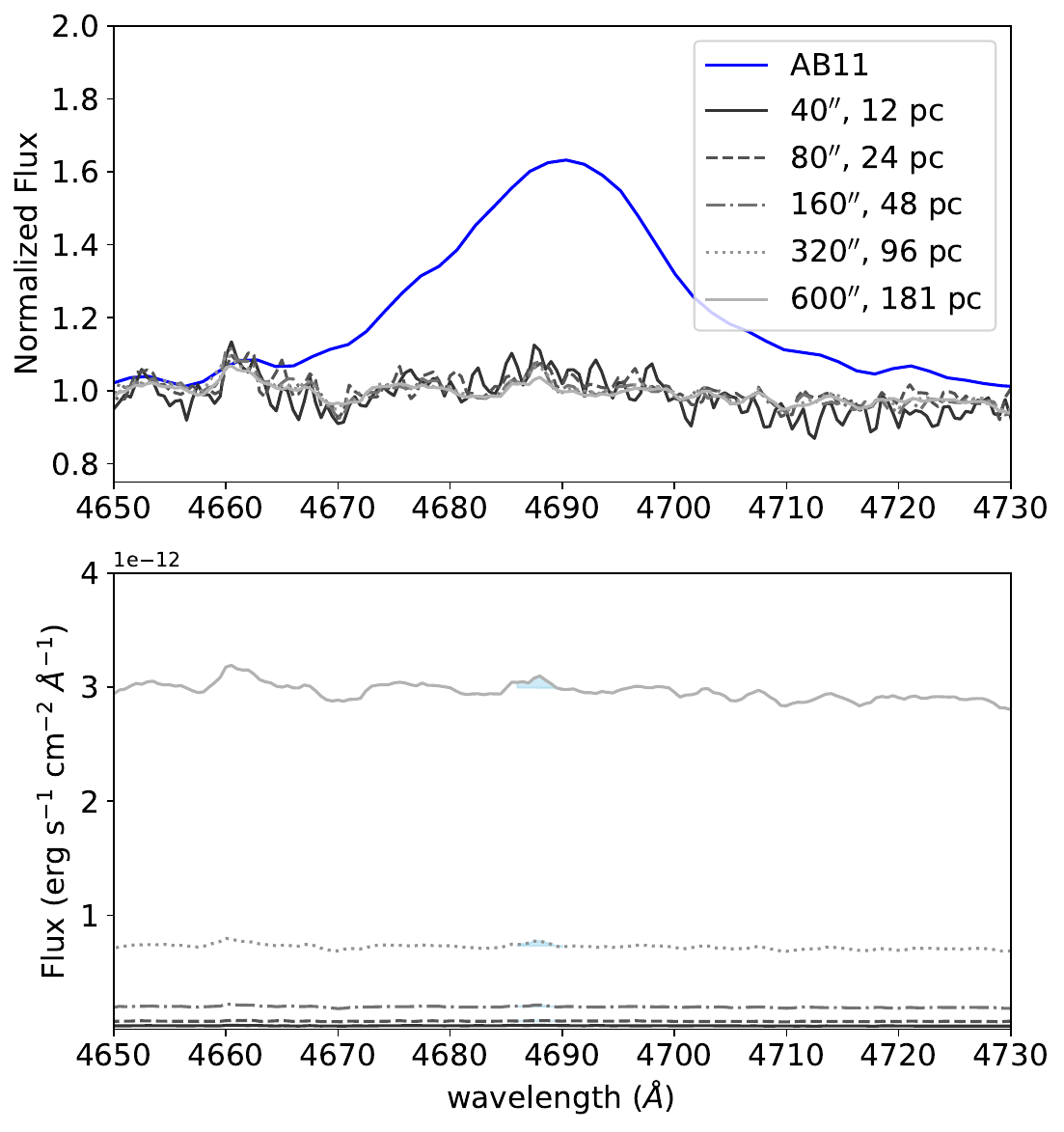}
      \caption{ Same as Fig.~\ref{fig:ab1heii} but for SMC AB11. 
      }
      \label{fig:ab11heii}
  \end{figure}

\begin{figure}[h!]
      \centering
      \includegraphics[width=1.\linewidth]{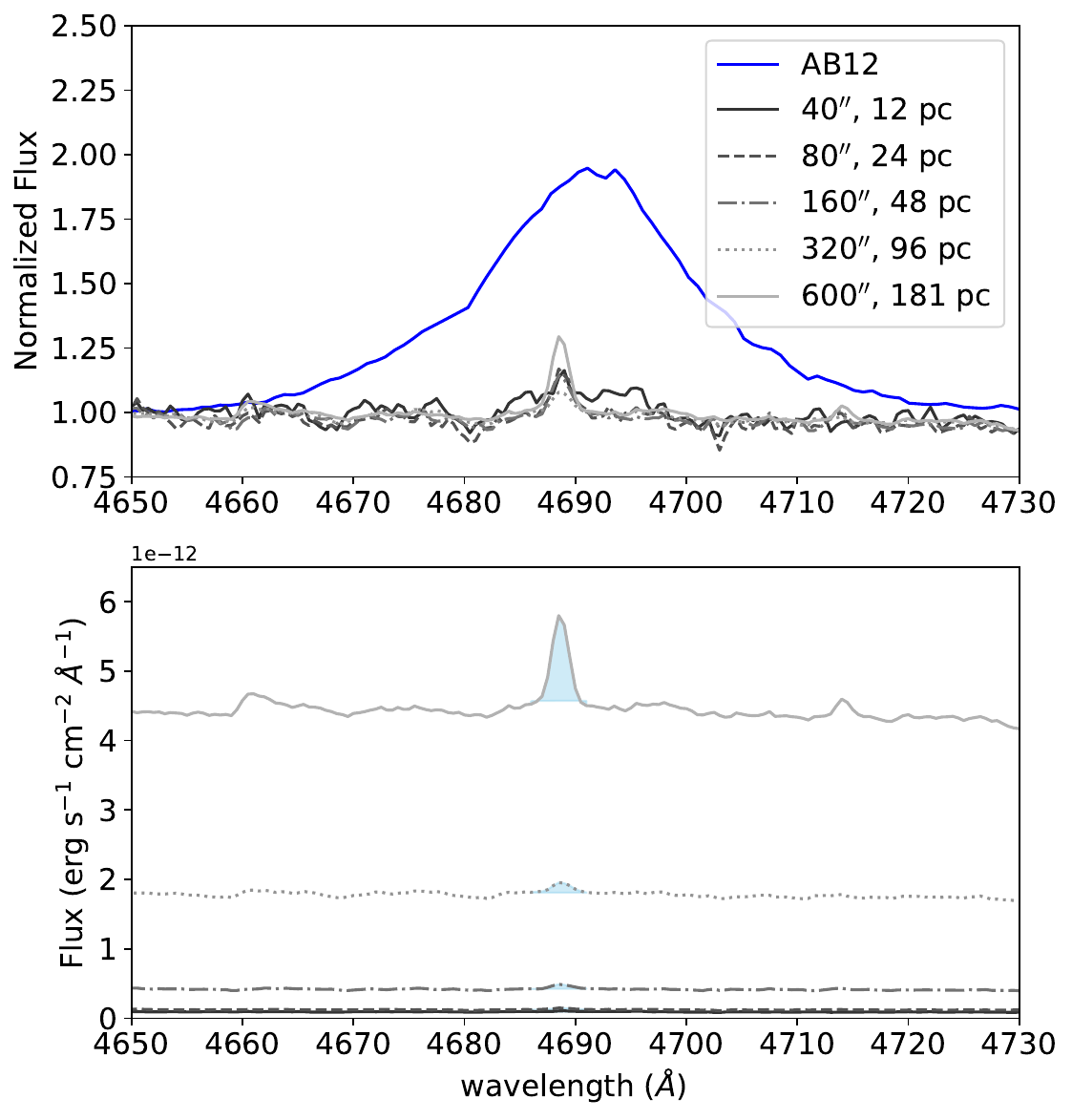}
      \caption{ Same as Fig.~\ref{fig:ab1heii} but for SMC AB12. 
      }
      \label{fig:ab12heii}
  \end{figure}

 Normalized and flux calibrated spectra around the \ion{He}{ii} $ - 4686\,$\AA\ line is shown in Figures.~\ref{fig:ab1heii} -- \ref{fig:ab12heii}, revealing notable differences between the targets:

\begin{itemize}
    \item SMC AB1: The small aperture of 40$\arcsec$ still shows some stellar \ion{He}{ii} contribution. We have fitted a gaussian distribution and detected faint present nebular emission, as seen in the lower panel of Fig.~\ref{fig:ab1heii}. The aperture of 320$\arcsec$ shows very weak and narrow \ion{He}{ii} nebular emission, while it is almost non-detectable for the 600$\arcsec$ aperture.
    \item SMC AB7: For the smaller aperture of 40$\arcsec$, the broad stellar component can still be detected, along with a strong nebular emission on top. For higher apertures only the narrow nebular emission is present. \citep{Tarantino+2024} has modeled the surrounding ISM gas for Spitzer IRS data using parameters from \citet{Shenar2016}, reproducing several ionizing lines in the far infrared region.
    \item SMC AB9: Similar to AB7, but weaker nebular emission. 
    \item SMC AB10: The nebular emission is very strong already in the 40$\arcsec$ aperture, while the broad stellar contribution is completely hidden.
    \item SMC AB11: Very faint to no emission is detected.
    \item SMC AB12: Faint stellar stellar emission in the 40$\arcsec$ aperture. The nebular emission increases with wider apertures. 
\end{itemize}

As only the area under the curves is of interest for this work, we did not perform a radial velocity correction for the LVM data.

\section{Detecting the blue WR bump}\label{ap:wrbump}
To get an estimate of the total number of WR stars in an integrated population showing WR bumps, \citet{Lopez-Sanchez2010} created a widely used formula based on the line luminosities provided by \citet{CrowtherHadfield2006} with a linear interpolation between the data for $Z_\odot$ and $Z_\odot /50$. However, these relations attribute the blue and red bumps only to late-type WN (WNL) and early-type WC stars and thus a WR population with mainly early-type WNE stars like in the SMC would be misinterpreted. In fact, any application of the \citet{Lopez-Sanchez2010} on the LVM SMC WNE data would yield numbers way below unity. 
More recently, \citet{Crowther+2023} presented detailed line luminosities for a large sample of WR stars from the Milky Way, the LMC and the SMC. For the summarized luminosity of early-type WN stars, \citet{Crowther+2023} calculates $L_\text{WN2-5}(\ion{He}{ii} \lambda 4686)$ at the SMC metallicity to be $1.7\times10^{35}$ erg\,s$^{-1}$ (see their Table 2).

Table~\ref{tab:nwnl} shows the estimated number of WN stars in our LVM regions using the luminosity values from \citet{Lopez-Sanchez2010,CrowtherHadfield2006,Crowther+2023} compared to our LVM $L_{\mathrm{HeII,VLM}}$ calculations (using the broad stellar component in the LVM flux calibrated spectra). We have used the apertures with a signal being 5\% greater than the noise, which corresponds to a diameter of $\leq$24 pc, while the for wider apertures the wide stellar component was completely not detectable.

\begin{table}
\caption{Target, the corresponding region and the estimated number of WN stars for 1/5 Z$_{\odot}$.}
\label{tab:nwnl}
\small
\centering
\begin{tabular}{c c c c }
\hline \hline
Target & Region\tablefootmark{a} & N$_{\mathrm{L10}}$ (1/5 Z$_{\odot}$)\tablefootmark{b} & N$_{\mathrm{C23}}$ (1/5 Z$_{\odot}$)\tablefootmark{c} \\
\hline
AB1&40$\arcsec$&0.158&1.213\\
&80$\arcsec$&0.130&0.999\\
\hline
AB7&40$\arcsec$&0.269&2.063\\
&80$\arcsec$&0.232&1.776\\
\hline
AB9&40$\arcsec$&0.096&0.734\\
&80$\arcsec$&0.146&1.122\\
\hline
AB10&40$\arcsec$&0.024&0.183\\
&80$\arcsec$&0.043&0.333\\
\hline
AB11&40$\arcsec$&0.075&0.573\\
&80$\arcsec$&0.078&0.600\\
\hline
AB12&40$\arcsec$&0.157&1.203\\
&80$\arcsec$&0.105&0.803\\
\hline
\end{tabular}
\tablefoot{
\tablefootmark{a}{Using the apertures with a signal being 5\% greater than the noise.}\tablefoottext{b}{Using Eq. 2 from \citet{Lopez-Sanchez2010} with the luminosities from Eq. 7 in \citet{Lopez-Sanchez2010} extrapolating to SMC metallicity.}
\tablefootmark{c}{From \citet{Crowther+2023}.}}
\end{table}

\section{EW ratios with respect to the distance}\label{ap:ews}
  \begin{figure}[h!]
      \centering
      \includegraphics[width=.8\linewidth]{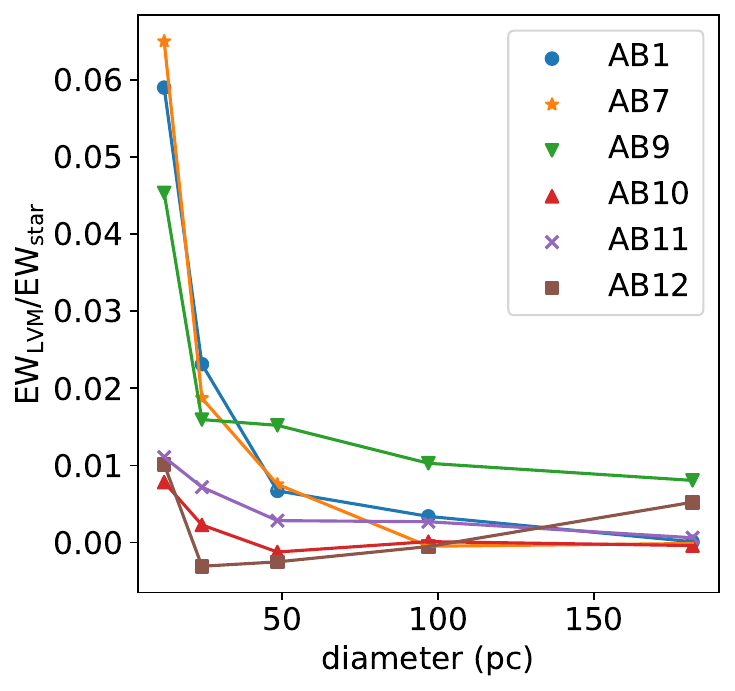}
      \caption{Ratio of Equivalent Widths (EW) of the extended stellar emission component of \ion{He}{ii}\,4686\,\AA\ seen by LVM to the intrinsic stellar EW. 
      }
      \label{fig:ews}
  \end{figure}

\end{document}